\documentclass{moriond}

\usepackage{amsmath,amssymb}

\graphicspath{{/home/archis/Dropbox/Research/MyPresentations/Figs/}}
\DeclareGraphicsExtensions{.pdf,.png,.jpg,.jpeg,.eps}

\newcommand{\Photo}{\includegraphics[height=35mm]{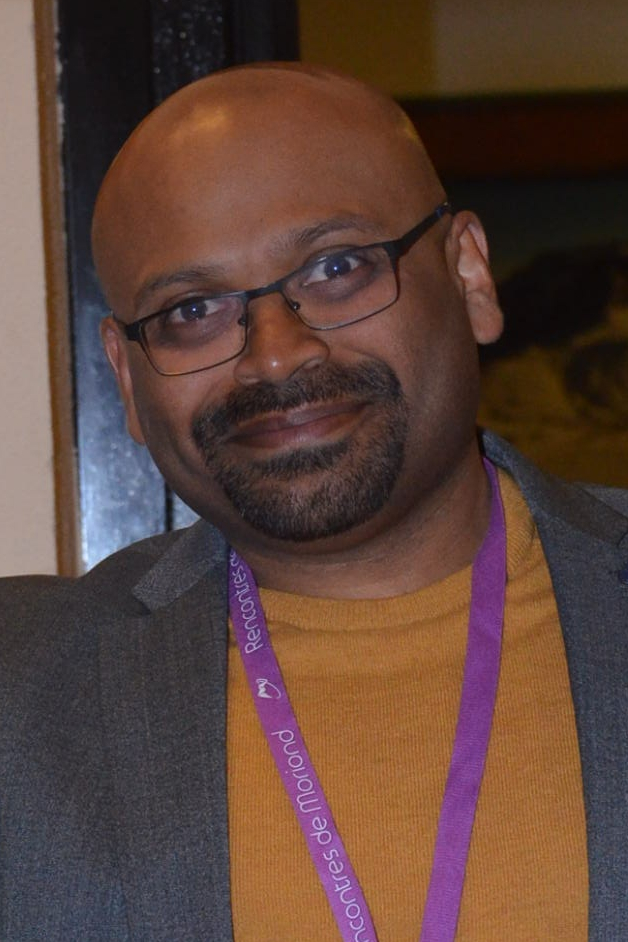}}

\begin{document}
\vspace*{4cm}
\title{Summary of cosmology with gravitational waves from compact binary coalescences}

\author{Archisman Ghosh \\ (for the LIGO Scientific Collaboration and Virgo Collaboration)}

\address{Nikhef -- National Institute for Subatomic Physics, \\ Science Park 105, 1098 XG Amsterdam, The Netherlands}

\maketitle\abstracts{
GW170817 with its coincident optical counterpart has led to a first ``standard siren'' measurement of the Hubble constant independent of the cosmological distance ladder~\cite{Abbott:2017xzu}. The Schutz ``statistical'' method~\cite{Schutz:1986gp}, which is expected to work in the absence of uniquely identified hosts, has also started bringing in its first estimates~\cite{Soares-Santos:2019irc}. In this work we report the current results of the gravitational-wave measurement of the Hubble constant and discuss the prospects with observations during the upcoming runs of the Advanced LIGO-Virgo detector network.}

\section{Introduction}

Gravitational-wave (GW) observations of coalescing compact binaries give us direct access to their luminosity distance making them standard distance indicators or ``standard sirens''~\cite{Schutz:1986gp,Holz:2005df,MacLeod:2007jd,Nissanke:2009kt}. With redshift from observed electromagnetic (EM) counterparts or identified host galaxies, they can be used to obtain a redshift-distance relationship and measure cosmological expansion and acceleration parameters. Of particular interest in context of present and upcoming detections by the Advanced LIGO and Virgo detector network is the Hubble constant $H_0$ -- the local rate of expansion of the universe at the present epoch. State-of-the-art measurements of $H_0$ coming from two complementary regimes, {\em i.e.}~by using local Type Ia supernovae and from the early universe cosmic microwave background, are in increasing tension with each other, the current discrepancy being at the $4.4$-$\sigma$ level~\cite{Aghanim:2018eyx,Riess:2019cxk}. An independent GW measurement of $H_0$ is thus consequential.

\section{$H_0$ with GW170817}

\begin{figure}[h]
\centering
\includegraphics[width=0.45\textwidth]{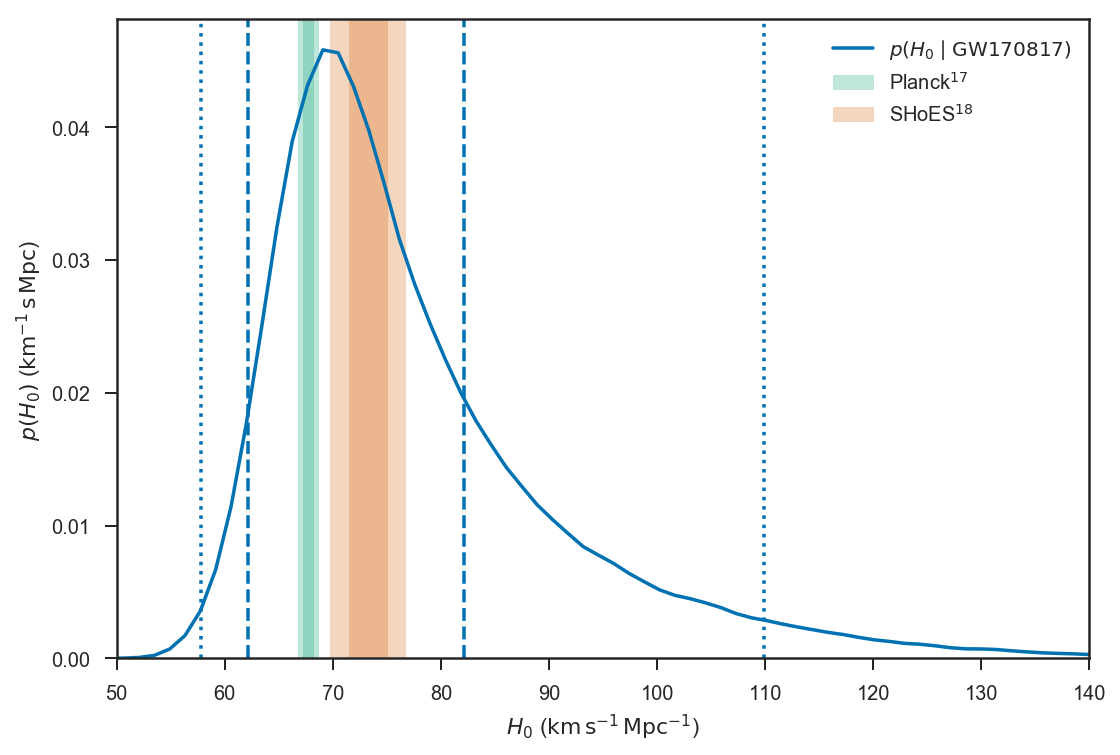}
\includegraphics[width=0.45\textwidth]{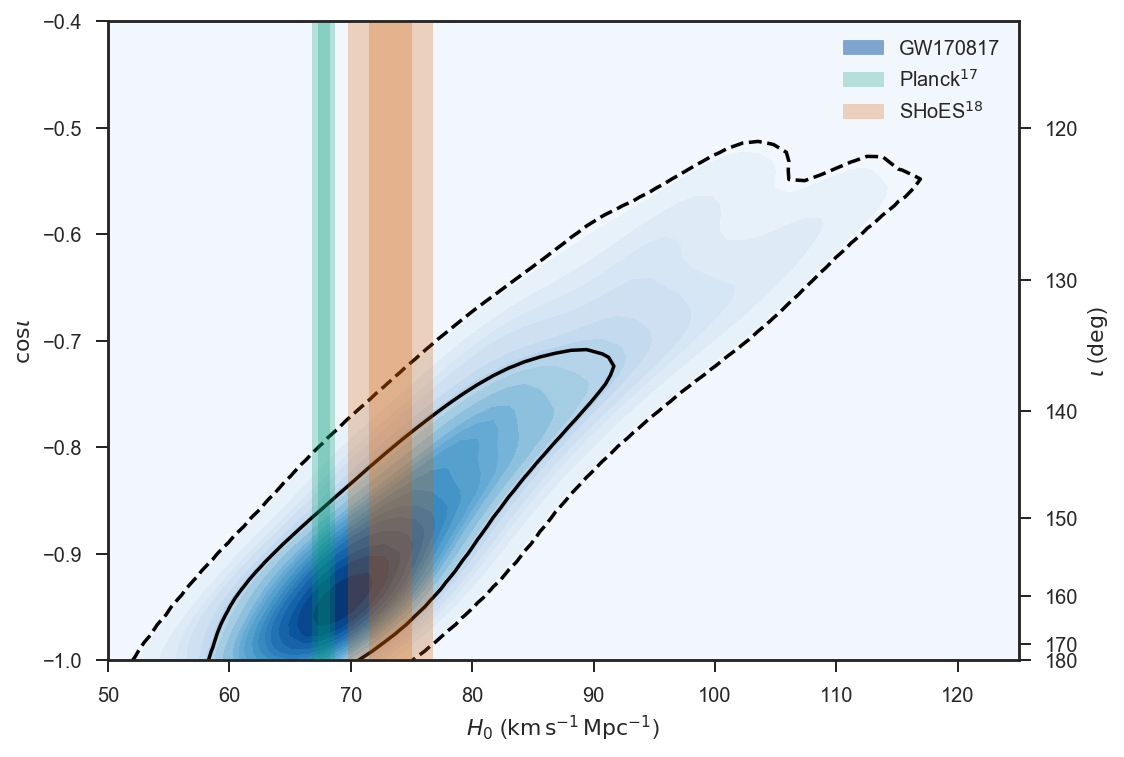}
\label{fig:H0}
\caption{Measurement of $H_0$ from GW170817. {\em Left panel:} The marginalized posterior density for $H_0$ (blue), with minimal $1$- and $2$-$\sigma$ ($68.3\%$ and $95.4\%$) credible intervals indicated respectively by dashed and dotted lines. {\em Right panel:} The joint posterior density on $H_0$ and inclination angle $\iota$ (blue) with $1$- and $2$-$\sigma$ contours indicated by solid and dashed (black) lines. Both panels additionally show constraints on $H_0$ from the cosmic microwave background (Planck) and supernovae (SH0ES). Plots are reproduced from the paper reporting the measurement.} 
\end{figure}

The binary neutron star (BNS) merger GW170817~\cite{TheLIGOScientific:2017qsa}, with its transient optical counterpart associated unambiguously with host galaxy NGC 4993, has given us the first standard siren measurement of $H_0$~\cite{Abbott:2017xzu}. In the local universe (distances $d_L\lesssim50\,\text{Mpc}$), the redshift-distance relationship is approximated to $H_0\,d_L \approx v_H$, where $v_H$ is the ``Hubble'' velocity, related to the cosmological redshift $z$ (and the speed of light $c$) by $v_H\equiv z\,c$. In order to obtain the Hubble velocity, the peculiar velocity of the host with respect to the Hubble flow needs to be estimated and subtracted from the observed recession velocity obtained directly from the spectral redshift. The Hubble velocity at the location of NCG 4993 is inferred to be $v_H=3,017\pm166\,\text{km}\,\text{s}^{-1}$. Together with samples of the GW distance posterior density (with $d_L=43.8^{+2.9}_{-6.9}\,\text{Mpc}$), the posterior density on $H_0$ shown on the left panel of Fig.~\ref{fig:H0} is obtained. This corresponds to $H_0=70.0^{+12.0}_{-8.0}\,\text{km}\,\text{s}^{-1}\text{Mpc}^{-1}$. With a fractional uncertainty of $\mathcal{O}(14\%)$, this measurement is broadly consistent with the state-of-the-art estimates, also shown on Fig.~2. More precise measurements will follow from subsequent observations. Since precision scales as $1/ \sqrt N$, where $N$ is the number of detections, $\mathcal{O}(200)$ similar detections are expected to take us to percent accuracy on the measurement of $H_0$~\cite{Nissanke:2013fka,Chen:2017rfc,Feeney:2018mkj,Mortlock:2018azx}. One of the main sources of uncertainties in the results is the correlation between GW distance and the inclination or viewing angle of the binary -- the GW amplitude from by a distant binary viewed face-on (or face-off) can be similar to that of a closer binary viewed edge-on. Choosing to not marginalize over the inclination angle $\iota$, one obtains a posterior density on the $H_0$-$\cos\iota$ plane as on the right panel of Fig.~\ref{fig:H0}. With independent information of the inclination angle, this can be used to refine the measurement of $H_0$. The very long-baseline interferometry (VLBI) observations of a relativistic jet along with the associated jet models~\cite{Mooley:2018dlz}, for example, have given a stronger estimate of the viewing angle of GW170817, leading to an improved $H_0$-estimate~\cite{Hotokezaka:2018dfi}.

\section{The ``statistical'' method using galaxy catalogues}

Optical counterparts will not be observed for all BNS mergers, and a counterpart might not always imply an unambiguous redshift. Furthermore, binary black holes (BBH) mergers, numerous of which are being detected, are not expected to have associated transient counterparts. In the absence of a uniquely identified host galaxy, one can use the method outlined by Schutz~\cite{Schutz:1986gp}. With sets of potential hosts in galaxy catalogues identified using the GW sky-localization, one can build up ``statistical'' information from multiple detections. While for a single detection in absence of a unique redshift from an identified host galaxy, one would likely obtain a multimodal posterior density for a cosmological parameter like $H_0$, with several detections one would be driven to the true value of the parameter, as has been demonstrated on simulations performed~\cite{DelPozzo:2011yh,Chen:2017rfc,Nair:2018ign}. With simulations performed on GW ``injections'' within galaxy catalogues assumed to be complete, it was shown that a $5\%$ estimate on $H_0$ can be obtained from $\mathcal{O}(100)$ sources without counterparts at $z\lesssim0.05$. In context of $H_0$ from upcoming observations, the ``statistical'' contribution is expected to come mainly from well-localized detections ($3$-D volume $\lesssim 10,000\,\text{Mpc}^3$), potentially leading to a $\mathcal{O}(10\%)$ precision in $H_0$ from only BBHs by 2026~\cite{Chen:2017rfc}. 

\begin{figure}[h]
\centering
\includegraphics[width=0.45\textwidth]{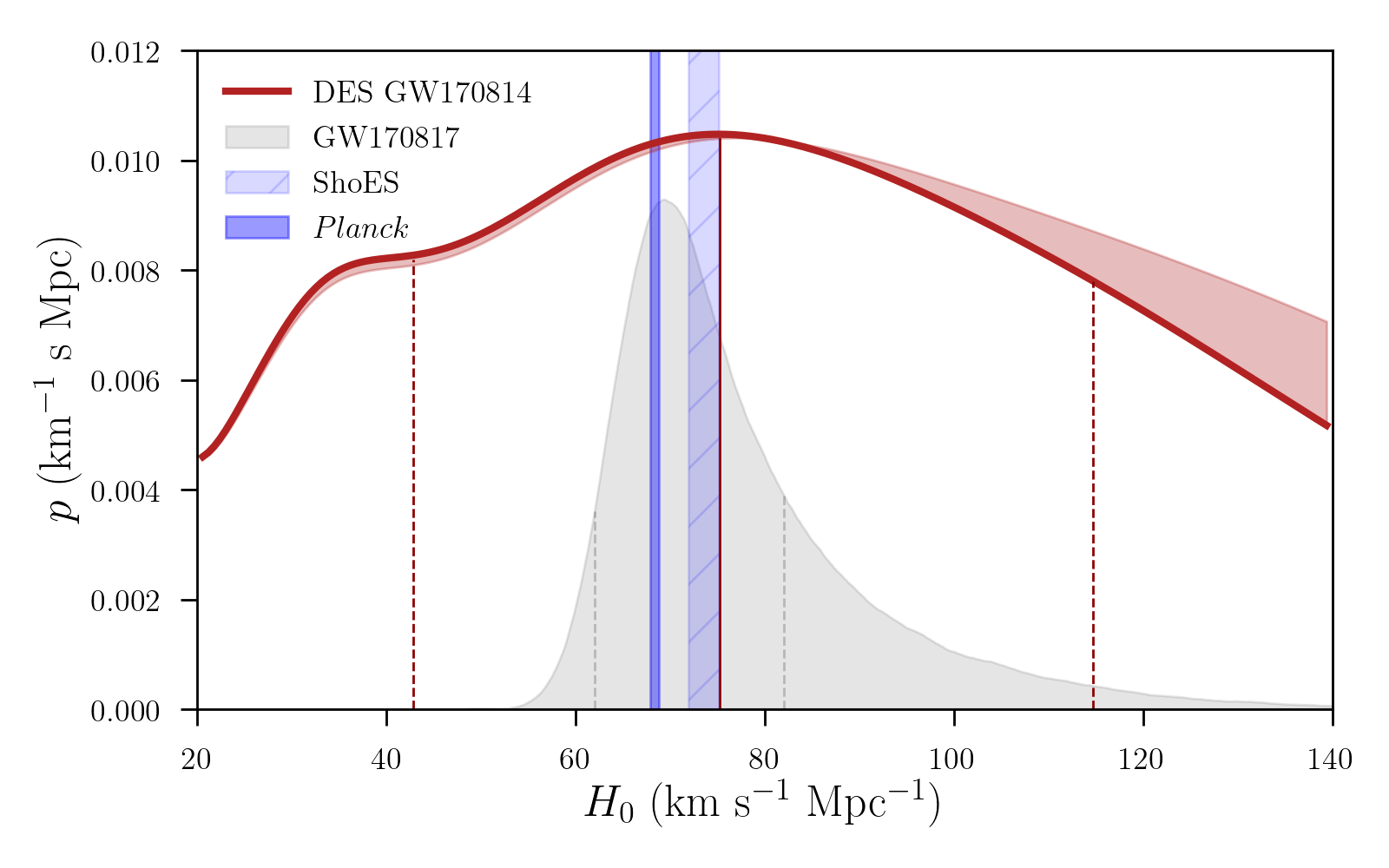}
\caption{Measurement of $H_0$ from GW170814 and possible host galaxies from the Dark Energy Survey. Host galaxies in the $90\%$ distance interval are used for the main result (solid red); dependence of the result on inclusion of more galaxies (up to $99.7\%$ of the distance interval) is indicated as a shaded band. $1$-$\sigma$ ($68\%$) credible intervals are indicated by dashed lines. GW170817 result (gray) and the Planck and Sh0ES measurements are included for comparison. The plot is reproduced from the paper reporting the measurement.} 
\label{fig:DES}
\end{figure}

The method of measuring $H_0$ using galaxy catalogues was illustrated for GW170817 assuming that no counterpart was observed~\cite{Fishbach:2018gjp}. A first galaxy-catalogue measurement of $H_0$ has now been performed, using the relatively well-localized BBH merger GW170814~\cite{Abbott:2017oio} (sky area $\approx 60\,\text{deg}^2$) and $\sim 77,000$ galaxies from the Dark Energy Survey (DES) which thoroughly followed up the GW170814 sky localization region~\cite{Soares-Santos:2019irc} (Fig.~\ref{fig:DES}). These results pave the path towards a more precise measurement of $H_0$ jointly with galaxy catalogues and GW data over the coming years.

\section{Selection effects}

A measurement of $H_0$ from multiple observations requires a precise understanding of selection effects; the results would be affected by a systematic bias if selection effects are not properly accounted for. GW selection effects (which arise due to a limited sensitivity of the GW detectors) are circumvented by dividing the ``biased'' result by a selection function or a detection efficiency obtained by integrating over all detectable data sets~\cite{Abbott:2017xzu,Chen:2017rfc,Mandel:2018mve,MDCpaper}. For the ``statistical'' method using galaxy catalogues, EM selection effects also become important: EM surveys are limited by the sensitivity threshold of telescopes and galaxy catalogues are ``incomplete''. The possibility that the GW host is not in the catalogue hence needs to be taken into account. This is done by marginalizing over the cases $g=G,\bar{G}$ where the host galaxy is respectively present or absent in the catalogue~\cite{MDCpaper}. The likelihood of the associated GW data $x_\text{GW}$ given a detection $D_\text{GW}$ and $H_0$ splits then into two terms:
\begin{align}
p(x_\text{GW}|D_\text{GW},H_0) &= \sum_{g=G,\bar{G}} p(x_\text{GW}|g,D_\text{GW},H_0)\,p(g|D_\text{GW},H_0)
\label{Eq:GbarG}
\end{align}
We finally have the in-catalogue and out-of-catalogue likelihood terms and the respective probabilities. The in-catalogue likelihood term is evaluated with information from the galaxy catalogue, namely, the galaxy redshifts $\{z_\text{gal}\}$ and sky coordinates, $\{\Omega_\text{gal}\}$.
\begin{align}
p(x_\text{GW}|G,D_\text{GW},H_0) &= \frac{1}{p(D_\text{GW}|G,H_0)} \sum_\text{gal} p(x_\text{GW}|z_\text{gal},\Omega_\text{gal},H_0)\,.
\label{Eq:incat}
\end{align}
The out-of-catalogue likelihood term is evaluated with a model assuming an apparent magnitude threshold $m_\text{th}$ for the threshold-limited survey, and priors on redshift sky-location and absolute magnitude distributions for galaxies $p(z,\Omega,M|H_0)$:
\begin{align}
p(x_\text{GW}|G,D_\text{GW},H_0) &= \frac{1}{p(D_\text{GW}|\bar{G},H_0)} \int_{z=z(m_\text{th},M,H_0)}^\infty dz\,d\Omega\,dM\,p(x_\text{GW}|z,\Omega,H_0)\,p(z,\Omega,M|H_0)\,.
\label{Eq:outcat}
\end{align}
The terms in the denominators of the above expressions are by evaluated integrating over all detectable data sets (in order to take into account the GW selection effects). Details of the full method to take into account selection effects are presented in a ``mock data challenge'' paper~\cite{MDCpaper}.

\section{A mock data challenge}

\begin{figure}[h]
\centering
\includegraphics[width=0.45\textwidth]{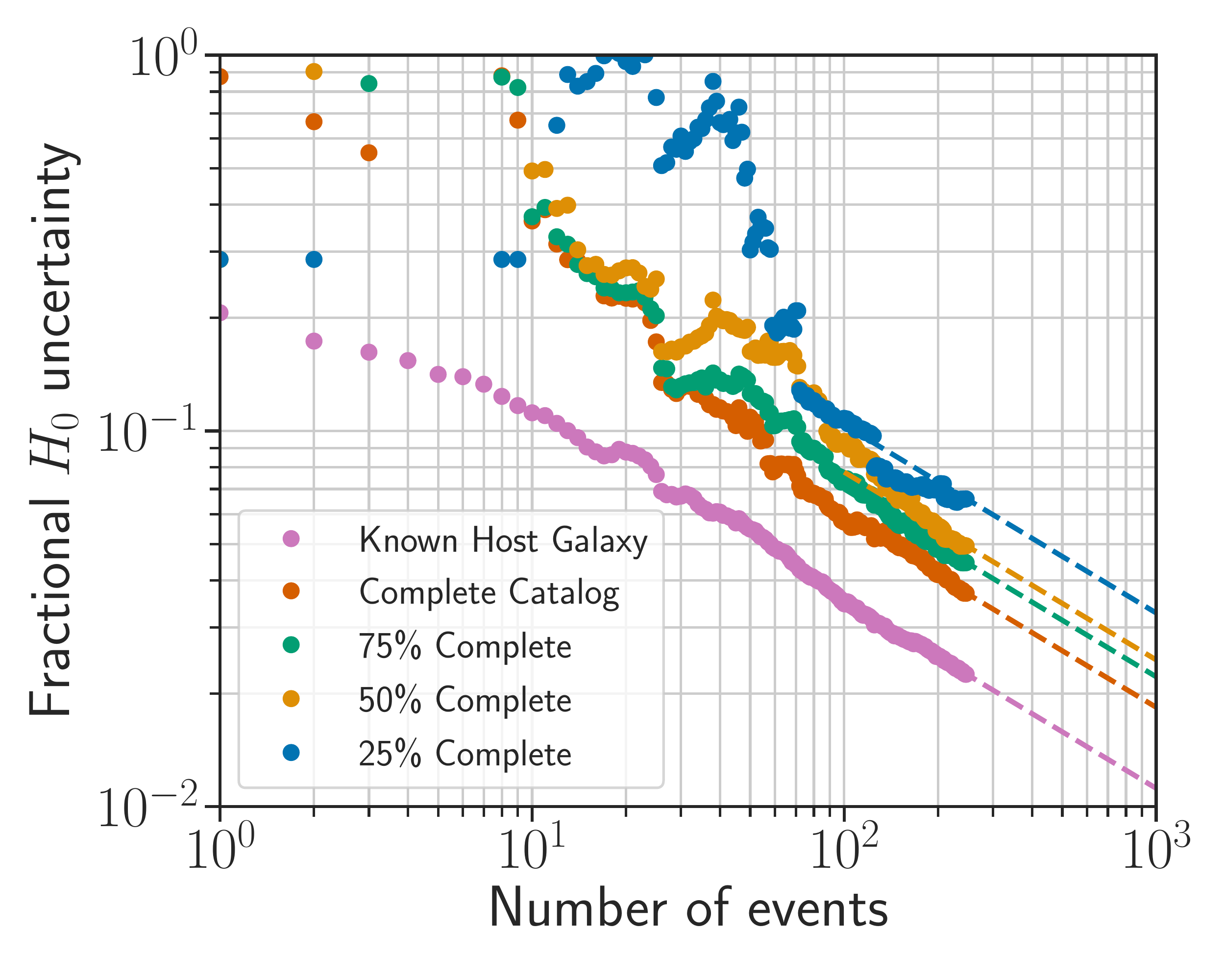}
\includegraphics[width=0.45\textwidth]{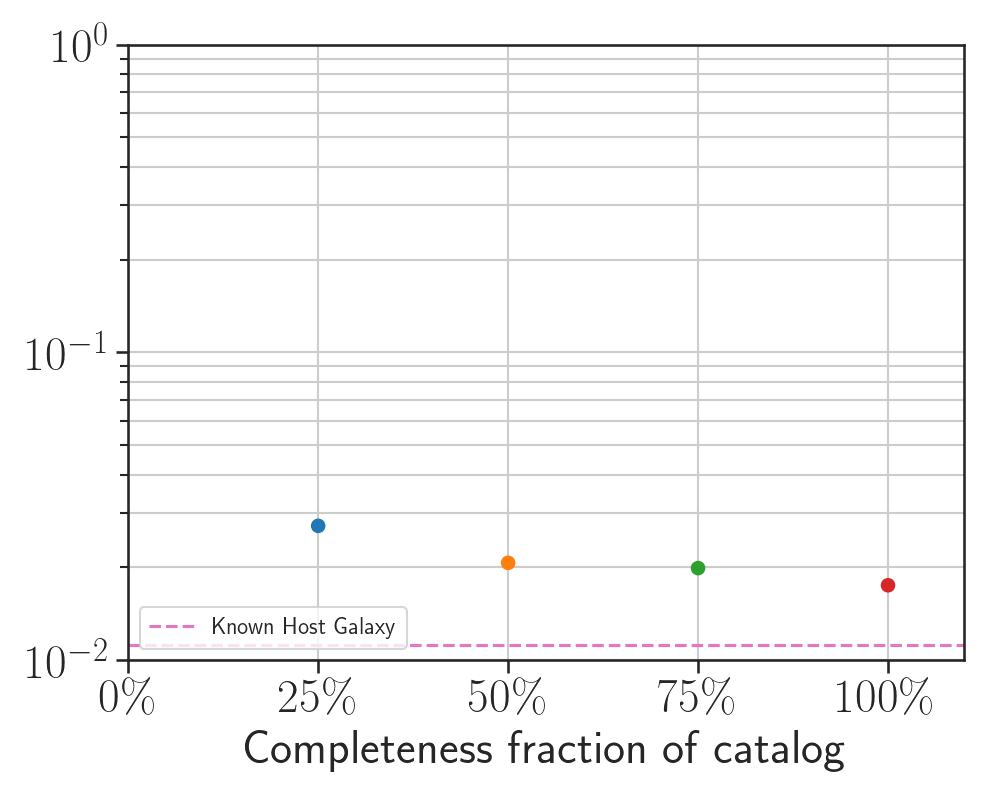}
\label{fig:MDC}
\caption{Results on the mock data challenge. {\em Left panel:} Fractional uncertainty on $H_0$ as a function of number of simulated events using galaxy catalogues of varying completeness. The ``counterpart'' result with a known host is also shown for comparison. {\em Right panel:} Fractional uncertainty on $H_0$ as a function of completeness of the galaxy catalogue for $\mathcal{O}(250)$ simulated events. Results are reproduced from the ``mock data challenge'' paper.} 
\end{figure}

The formalism presented above has been tested on a series of simulations~\cite{MDCpaper}. Simulated data from $\mathcal{O}(250)$ BNS events~\cite{Singer:2014qca} have been chosen in association with simulated galaxy catalogues~\footnote{The simulated galaxy catalogues are about 70 times sparse compared to real galaxy catalogues.} of varying completeness. The results are shown in Fig.~\ref{fig:MDC}. They show a convergence as $1/\sqrt{N}$ with the number of detections $N$. For these simulations, depending on the completeness of the catalogue, the precision reached with the ``statistical'' are only a few {\em i.e.}~$\mathcal{O}(2$-$5)$ times broader than the ``counterpart'' case where the host galaxy is assumed to be known.

\section{Summary and outlook}

GWs from compact binaries provide an measurement of $H_0$ {\em independent} of other existing estimates. Numerous detections over the coming years, with or without counterparts, will lead to a precise GW-measurement of $H_0$, competing with its currently conflicting state-of-the-art estimates. In order to get an unbiased result, it is crucial to account for any possible systematic effects, the latter becoming comparable to the smaller statistical uncertainties with an increasing number of detections. Peculiar velocities are comparatively large at low redshifts, and need to be estimated and accounted for. Selection effects are of prime importance. To account for GW selection effects one needs to integrate over all detectable data sets, which requires a marginalization over all GW parameters that characterize the detectable data. With galaxy catalogues, it becomes important to account for selection effects coming from threshold-limited EM surveys, and to characterize statistical and systematic uncertainties associated with redshift and luminosity measurements. Effects like GW lensing are going to become important further down the line. A better quantification of uncertainties associated with GW detector calibration will also become imminent moving towards percent-precision.

\section*{Acknowledgments}

The authors gratefully acknowledge the support of the United States
National Science Foundation (NSF) for the construction and operation of the
LIGO Laboratory and Advanced LIGO as well as the Science and Technology Facilities Council (STFC) of the
United Kingdom, the Max-Planck-Society (MPS), and the State of
Niedersachsen/Germany for support of the construction of Advanced LIGO 
and construction and operation of the GEO600 detector. 
Additional support for Advanced LIGO was provided by the Australian Research Council.
The authors gratefully acknowledge the Italian Istituto Nazionale di Fisica Nucleare (INFN),  
the French Centre National de la Recherche Scientifique (CNRS) and
the Foundation for Fundamental Research on Matter supported by the Netherlands Organisation for Scientific Research, 
for the construction and operation of the Virgo detector
and the creation and support  of the EGO consortium. 
The authors also gratefully acknowledge research support from these agencies as well as by 
the Council of Scientific and Industrial Research of India, 
the Department of Science and Technology, India,
the Science \& Engineering Research Board (SERB), India,
the Ministry of Human Resource Development, India,
the Spanish  Agencia Estatal de Investigaci\'on,
the Vicepresid\`encia i Conselleria d'Innovaci\'o, Recerca i Turisme and the Conselleria d'Educaci\'o i Universitat del Govern de les Illes Balears,
the Conselleria d'Educaci\'o, Investigaci\'o, Cultura i Esport de la Generalitat Valenciana,
the National Science Centre of Poland,
the Swiss National Science Foundation (SNSF),
the Russian Foundation for Basic Research, 
the Russian Science Foundation,
the European Commission,
the European Regional Development Funds (ERDF),
the Royal Society, 
the Scottish Funding Council, 
the Scottish Universities Physics Alliance, 
the Hungarian Scientific Research Fund (OTKA),
the Lyon Institute of Origins (LIO),
the Paris \^{I}le-de-France Region, 
the National Research, Development and Innovation Office Hungary (NKFIH), 
the National Research Foundation of Korea,
Industry Canada and the Province of Ontario through the Ministry of Economic Development and Innovation, 
the Natural Science and Engineering Research Council Canada,
the Canadian Institute for Advanced Research,
the Brazilian Ministry of Science, Technology, Innovations, and Communications,
the International Center for Theoretical Physics South American Institute for Fundamental Research (ICTP-SAIFR), 
the Research Grants Council of Hong Kong,
the National Natural Science Foundation of China (NSFC),
the Leverhulme Trust, 
the Research Corporation, 
the Ministry of Science and Technology (MOST), Taiwan
and
the Kavli Foundation.
The authors gratefully acknowledge the support of the NSF, STFC, MPS, INFN, CNRS and the
State of Niedersachsen/Germany for provision of computational resources.
This document is LIGO-P1900146-v2.

\section*{References}

\bibliography{inspirerefs}

\end{document}